\documentclass{ieeeaccess}
\renewcommand{\baselinestretch}{0.95}
\usepackage{cite}
\usepackage{amsmath,amssymb,amsfonts}
\usepackage{algorithm}
\usepackage{graphicx} 
\usepackage{textcomp} 
\usepackage{comment}
\usepackage{array}
\newcolumntype{P}[1]{>{\centering\arraybackslash}p{#1}}
\usepackage{listings}
\usepackage{algpseudocode}
\usepackage{csquotes}
\usepackage{eso-pic}
\usepackage{float} 
\usepackage{subfigure}
\usepackage{booktabs}
\usepackage{url}
\usepackage{makecell}
\usepackage{hyperref}
\usepackage{multirow}
\usepackage{tabularx,booktabs}
\newcolumntype{C}{>{\centering\arraybackslash}X}

\makeatletter
\newcommand*{\algrule}[1][\algorithmicindent]{\makebox[#1][l]{\hspace*{.5em}\vrule height .75\baselineskip depth .25\baselineskip}}%

\newcount\ALG@printindent@tempcnta
\def\ALG@printindent{%
    \ifnum \theALG@nested>0
        \ifx\ALG@text\ALG@x@notext
            \addvspace{-3pt}
        \else
            \unskip
            \ALG@printindent@tempcnta=1
            \loop
                \algrule[\csname ALG@ind@\the\ALG@printindent@tempcnta\endcsname]%
                \advance \ALG@printindent@tempcnta 1
            \ifnum \ALG@printindent@tempcnta<\numexpr\theALG@nested+1\relax
            \repeat
        \fi
    \fi
    }%
\usepackage{etoolbox}
\patchcmd{\ALG@doentity}{\noindent\hskip\ALG@tlm}{\ALG@printindent}{}{\errmessage{failed to patch}}
\makeatother

\def\BibTeX{{\rm B\kern-.05em{\sc i\kern-.025em b}\kern-.08em
    T\kern-.1667em\lower.7ex\hbox{E}\kern-.125emX}}

\begin{document}

\history{Date of publication December 00, 2020, date of current version December 00, 2021.}
\doi{10.1109/ACCESS.2020.DOI}

\title{Enhancing Data Security for Cloud Computing Applications through Distributed Blockchain-based SDN Architecture in IoT Networks}

\author{\uppercase{Anichur Rahman\authorrefmark{1, 2}},
\uppercase{Md. Jahidul Islam}\authorrefmark{3},
\uppercase{Rafiqul Islam}\authorrefmark{4},
\uppercase{Dipanjali Kundu}\authorrefmark{1},
\uppercase{Md. Razaul Karim}\authorrefmark{2},
\uppercase{Ziaur Rahman}\authorrefmark{5},
\uppercase{Shahab S. Band}\authorrefmark{6}
}

\address[1]{Department of Computer Science and Engineering, National Institute of Textile Engineering and Research (NITER), Constituent Institute of the University of Dhaka, Savar, Dhaka-1350, Bangladesh. (e-mail: anis.mbstu.cse@gmail.com, 
dipanjali\_kundu@niter.edu.bd 
)}

\address[2]{Department of Computer Science and Engineering, Mawlana Bhashani Science and Technology University, Tangail, Bangladesh (e-mail: razaulkarimce15004@gmail.com)
}

\address[3]{Department of Computer Science and Engineering, Green University of Bangladesh, Dhaka, Bangladesh (e-mail: jahidul.jnucse@gmail.com)}

\address[4]{School of Computing and Mathematics, Faculty of Business, Justice and Behavioural Sciences, Charles Sturt University, Australia (e-mail: mislam@csu.edu.au)}

\address[5]{Department of Information and Communication Technology, Mawlana Bhashani Science and Technology University, Tangail, Bangladesh (e-mail: zia@iut-dhaka.edu)}

%

%
\address[6]{Future Technology Research Center, College of Future, National Yunlin University of Science and Technology, Douliou, Taiwan, ROC (e-mail: shamshirbands@yuntech.edu.tw)}

%

%

\tfootnote{
}

\markboth
{A. Rahman \headeretal: Enhancing Data Security for Cloud Computing Applications...}
{A. Rahman \headeretal: Enhancing Data Security for Cloud Computing Applications...}

\corresp{Corresponding author:
Rafiqul Islam (e-mail: mislam@csu.edu.au); Shahab S. Band (e-mail: shamshirbands@yuntech.edu.tw); Anichur Rahman (e-mail: anis.mbstu.cse@gmail.com).}

\begin{abstract}
Blockchain (BC) and Software Defined Networking (SDN) are some of the most prominent emerging technologies in recent research. These technologies provide security, integrity, as well as confidentiality in their respective applications. Cloud computing has also been a popular comprehensive technology for several years. \textcolor{black}{Confidential information is often shared with the cloud infrastructure to give customers access to remote resources, such as computation and storage  operations.} However, cloud computing also presents substantial security threats, issues, and challenges. 
Therefore, to overcome these difficulties, we propose integrating Blockchain and SDN in the cloud computing platform. In this research, we introduce the \enquote{DistB-SDCloud} architecture to better secure clouds. Moreover, we leverage a distributed Blockchain approach to convey security, confidentiality, privacy, integrity, adaptability, and scalability in the proposed architecture. BC provides a distributed or decentralized and efficient environment for users. Also, we present an SDN approach to improving the reliability, stability, and load balancing capabilities of the cloud infrastructure. Finally, we provide an experimental evaluation of the performance of our SDN and BC-based implementation using different parameters, also monitoring some attacks in the system and proving its efficacy.
\end{abstract}


\titlepgskip=-15pt
\maketitle

\section{Introduction}
\label{sec:introduction}
\PARstart{I}{n} today's interconnected world, cloud technology is seen as a crucial enabler of IT industry innovation.
It is a model that provides consumers with various on-demand services, and network access to  shared databases of physical resources such as computation and storage.
In this way, customers do not have to buy expensive hardware anymore, they can access these services using commodity hardware (such as a laptop) connected to the Internet, giving them the means to develop solutions to complex problems. \textcolor{black}{Furthermore, cloud computing enables users to access resources from any location remotely, allowing for virtual collaboration. It enables users to improve resources relatively quickly; previously,this was time-demanding with traditional hardware-based computing systems. Proper resource usage aids in mitigating the over and under-utilization problem \cite{shukri2021enhanced}.}
Cloud computing provides a range of services like Software as a Service, Platform as a Service, and Infrastructure as a Service, which are in short known as SaaS, PaaS, IaaS \textcolor{black}{respectively}. Also,cloud services are scalable, flexible, and reliable on-demand for users \cite{9350419, dillon2010cloud}.  
Another technology is Blockchain, but it has a  focus on security. For this reason, Blockchain technology has attracted particular interest in the financial environment, and in general wherever data security is of topmost priority. \textcolor{black}{Blockchain with the addition of secret sharing security enhancement is possible in  cloud services with the improvement of data security \cite{cha2021blockchain}.  Further, Blockchain technology helps detect malicious use by enforcing a distrustful provider to detect suspicious suppliers \cite{9350231}.} 

In this architecture, a hash value is generated to secure the information transmitted through the Blockchain structure. Further, Blockchain stores the information in a public ledger where any change can be noticed by others connected to the ledger. This is why no third party can disturb the transaction \cite{park2017blockchain}. Furthermore, as cloud computing is interconnected with many security issues and financial transactions, Blockchain can strengthen the reliability of the cloud computing functionalities \cite{9499121}.  
Another emerging paradigm is SDN \cite{Rahman2020}, which makes network management is more straightforward.It has found extensive application in cloud computing, where networks are challenging to manage and troubleshoot.
As already stated, network performance is crucial to the quality of service of the final users, who access shared resources provided over the Internet. Another emerging technology is an SDN used to control a network employing some program. With the use of an SDN, it is easy to manage  network issues, as cloud computing is full of network issues. Without the network, cloud computing is nothing since the resources will be shared and provided to the users over the Internet. Recent applications of SDN employ multiple controllers, adding a new dimension to the network of devices. As the clients of this kind of cloud computing have seen a steady increase over recent years, the management of the resources and security of the data have become a primary concern. \textcolor{black}{To manage the resources efficiently SDN can be an effective solution that will trace the network's traffic and estimate the bandwidth of the network \cite{alomari2021resource}.}
The combination of new emerging technologies is crucial to protect the data and make the activity of cloud easier. \vspace{1mm}

From the above analysis, we propose a distributed, secure Blockchain-based SDN enabled control architecture for cloud computing. 
A distributed Blockchain provides reliable and efficient security privately and publicly in the cloud environment efficiently. The following are the paper's contributions: \vspace{1mm}

\begin{itemize}
  \item We introduce a distributed structure to enhance the reliability and speed of physical and logical data on the cloud infrastructure. \vspace{1mm}
  
  \item To improve the security, privacy, and secrecy of the presented architecture, we use a distributed SDN-Blockchain strategy. \vspace{1mm}  
  
  \item Finally, we investigate the feasibility of the given cloud model in terms of various characteristics and evaluate its responsiveness to network threats.
\end{itemize}  
\vspace{2mm}
\textcolor{black}{Some of the notations are listed in Table \ref{tab:aphaNotation}. The remainder of the paper is organized as follows:  The background and literature review of the current work are discussed in Section II. 
Further, a \enquote{DistB-SDCloud} architecture for cloud computing has been presented in Section III, and we also present this architecture in a differently. In Section IV, performance analysis and discussion are presented. In addition, we suggest some future scopes in Section V. Finally, the authors conclude this article in Section VI, outlining the limitations of this work and future research plans}.

\begin{table}[!h]
\caption{\textcolor{black}{List of acronyms in alphabetical order.}}
\label{tab:aphaNotation}
\centering
\resizebox{\columnwidth}{!}
{
\begin{tabular}{@{}p{1.5cm}p{6.5cm}@{}}
\toprule
\textbf{Notations} & \textbf{Description} \\
\midrule
\emph{AI} & Artificial Intelligence \\
\emph{API} & Application Programming Interface \\
\emph{BC} & Blockchain \\
\emph{BCF} & Blockchain Fundamental\\
\emph{BCT} & Blockchain Technology\\
\emph{CC} & Cloud Computing \\
\emph{DoS} & Denial of Service \\
\emph{DDoS} & Distributed Denial of Service \\
\emph{DS} & Data Security \\
\emph{DM} & Data Management \\
\emph{IoT} & Internet of Things \\
\emph{NFV} & Network Function Virtualization \\
\emph{PoW} & Proof of Work \\
\emph{QoS} & Quality of Service \\
\emph{SC} & Smart Contact \\
\emph{SDN} & Software Defined Networking \\
\emph{SaaS} & Software as a Service \\
\emph{PaaS} & Platform as a Service \\
\emph{IaaS} & Infrastructure as a Service \\
\emph{RTR} & Response to Request time \\
\emph{CBR} & Constant Bit Rate \\

\bottomrule
\end{tabular}
}
\end{table}

\section{Background and Literature Review}
\textcolor{black}{Recently, several researchers have completed many articles based on  leading emerging technologies like SDN, Blockchain, cloud computing, and other smart technological applications. In this section, we present a systematic literature overview of the current study based-on these technologies, }\vspace{2mm}

\subsection{Software Defined Networking (SDN)}
\begin{figure}[h]
    \centering
    \includegraphics[scale=0.39]{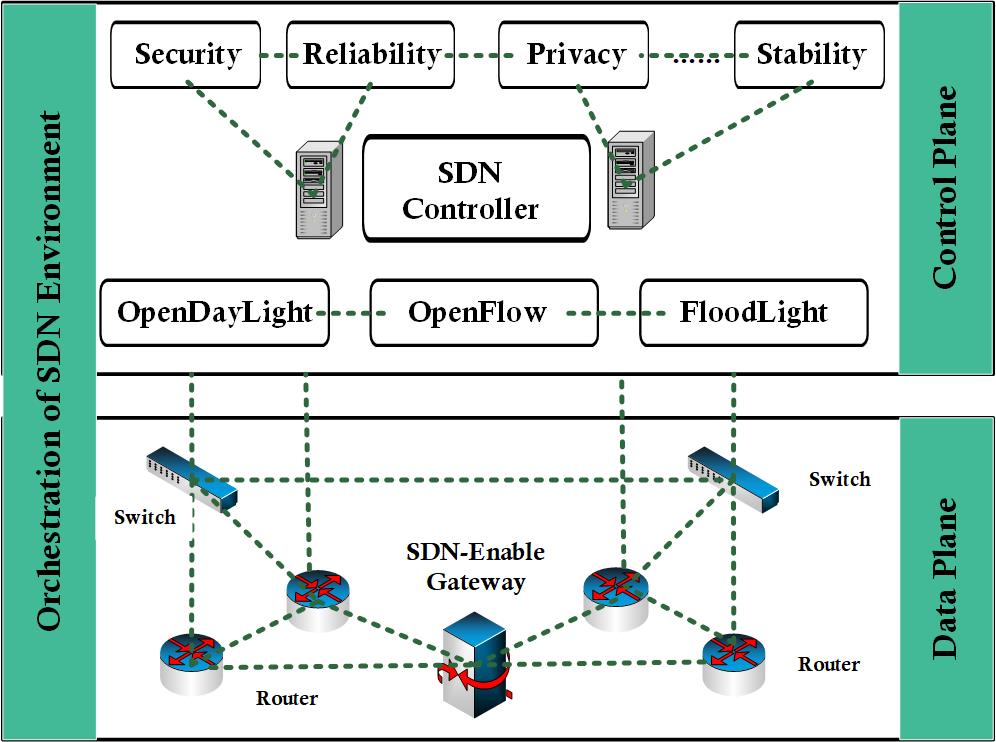}
   \caption{SDN architecture}
    \label{fig:sdna}
\end{figure}

\textcolor{black}{
Software Defined Networking is a networking paradigm, that helps manage and configure the network efficiently \cite{islam2019distblacknet}.
It can provide a logically centralized approach to managing data and network resources.
This orchestration and management strategy can be controlled by distinct and decoupled planes,as shown in Fig. \ref{fig:sdna}.
}

\textcolor{black}{The SDN gateway offers forwarding capabilities to IoT devices, allowing them to enter the SDN environment.}
Sahay et al. discussed the application of SDN for the security of computer networking, as the programmability of  SDN offers network security improvements \cite{sahay2019application, Islam_Rahman_Kabir_Khatun_Pritom_Chowdhury_2021}. The authors provided a broad, comprehensive review of SDN applied for detecting and mitigating   security issues. In a similar study, Shin et al. presented an overview of SDN based security in the networking system in  \cite{shin2016enhancing}. This survey also provided many opportunities for improving SDN features in different fields. Finally, the authors guided some futures research based on SDN security processes in different domains.

Moreover, machine learning and entropy-based mechanism were used to detect vulnerabilities.
Abdelaziz et al. suggested a new controller cluster to address the issues of software-defined network reliability, manageability, fault tolerance, and interoperability. \cite{abdelaziz2017distributed}. Clustering in the control plane can reduce transmission delay and packet loss significantly. The authors also claimed that their presented scheme optimizes the controller’s  performance by achieving reasonable CPU utilization. 
OpenFlow, OpenStack, and OpenDayLight are three existing SDN platforms that have been presented with an architecture \cite{chourishi2015role}. The authors also discussed load balancing and security materials for the SDN framework to lower the cost of related issues.

\subsection{Blockchain Technology (BCT)}

\begin{figure}[h]
    \centering
    \includegraphics[scale=0.38]{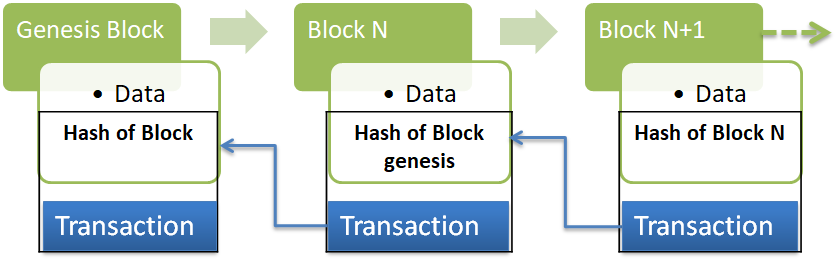}
   \caption{Blockchain procedure}
    \label{fig:bct}
\end{figure}

\textcolor{black}{
Blockchain is a decentralized/distributed ledger, that ignores the party involvement from  different transactions \cite{el2018decision}. Its work is completed correctly using consensus processes such as proof-of-work (PoW) and proof-of-stack (PoS).
Meanwhile, it uses hash data to maintain communication from one block to the next, as shown in Fig. \ref{fig:bct}. 
}
\textcolor{black}{Further, it can prevent information tampering and data hacking.
At present, this technology has been used in different fields like the IoT, healthcare, electronic voting, cloud computing, and so on \cite{wang2018secure}.}
In a similar paper, Koens et al. mentioned  Blockchain technology  and how to use it in a business setting \cite{koens2018Blockchain}. 
They aimed at answering two fundamental questions: first, is Blockchain  needed, and, if this is the case, which Blockchain is the right one .
Blockchain is often required to consider the results of this survey. The authors kept the Blockchain in addition to other database technologies. Finally, their schemes showed some opposition and suggest that no scheme is complete, and they do not describe the limitations of Blockchain technology. 
Treiblmaier et al. addressed a scenario of distributing Blockchain, including its future work  \cite{treiblmaier2019blockchain}. Meanwhile, the authors built a self determination theory (SDT), providing a guideline for readers on three different scenarios.

\vspace{2mm} 

\begin{table*}[h!]
\caption{Review of existing literature}
\vspace{2mm}
 \label{tab:related_work}
\centering
\begin{tabular}{ |p{2cm}|P{1.8cm} |P{1.5cm}| P{2.4cm}| P{2.7cm} |P{2.5cm}|}
\hline
\textbf{Authors} & \textbf{Key Technologies} & \textbf{Framework} & \textbf{Implementation} & \textbf{
Challenges Address} & \textbf{Blockchain Application Framework} \\
\hline

Sharma et al.  \cite{sharma2017distblocknet}  & Blockchain (BC), SDN & Centralized, Distributed & IoT Ecosystem & Reliability \& Privacy  & Ethereum Network \\ 
\hline

Rahman et al. \cite{9290627} & BC, SDN & Centralized, Distributed & Smart City & Reliability \& Privacy  & Ethereum Network \\ 
\hline

Liang et al. \cite{liang2017provchain}   & BC & Decentralized & Cloud Computing & Reliability \& Privacy  & Ethereum Network \\ \hline

Nguyen et al. \cite{nguyen2020integration} & BC & Decentralized & Smart Healthcare, Smart City, and Smart Industry & Reliability \& Privacy  & Ethereum \& Hyperledger Fabric \\ \hline

Sharma et al. \cite{sharma2017software}   & SDN & Centralized, Distributed & Edge Computing & Energy & - \\ 
\hline

Rahman et al. \cite{rahman2021sdn}   & IoT, BC, SDN & Centralized, Distributed & Cloud Computing & Energy \& Privacy & - \\ 
\hline

yang et al. \cite{yang2020authprivacychain}   & BC, Cloud Computing & Centralized, Distributed & Cloud Computing & Reliability \& Privacy  & - \\ 
\hline

Singh et al. \cite{singh2019sh} & BC, SDN & Centralized, Distributed & Smart City & Reliability \& Privacy & Ethereum Network \\ 
\hline

Gaetani et al. \cite{gaetani2017blockchain} & BC, Cloud Computing & Distributed & Utility Computing & Reliability \& Privacy & Ethereum \& Ethereum\\ 
\hline

Cech et al. \cite{cech2019fog} & BC, SDN & Distributed  & Fog Computing &  Reliability \& Privacy & Hyperledger Network \\ 
\hline

Li et al. \cite{li2018toward} & BC & Distributed & Cloud manufacturing system & Data Availability \& Privacy & Ethereum \\ 
\hline

Rehman et al. \cite{rehman2019cloud} & BC &  Distributed & Cloud Computing, Resource constrained devices & Reliability \& Privacy & Ethereum Network \\ 
\hline

Tuli et al. \cite{tuli2019fogbus} & BC & Distributed & IoT Networks & Reliability \& Privacy & - \\ 
\hline

Deep et al. \cite{deep2019authentication} & BC & Decentralized & Cloud Computing & Security \& Reliability & Ethereum \\ 
\hline

Park et al. \cite{park2017blockchain} & BC & - & IoT, Cloud Computing & Reliability \& Privacy & Ethereum \\ 
\hline

Wei et al. \cite{wei2020blockchain} & BC & Decentralized & Cloud Computing & Reliability \& Security & -\\ 
\hline

Hasan et al. \cite{9333523} & BC & Decentralized & Verification System & Reliability \& Privacy & Ethereum\\ 
\hline

\end{tabular}
\end{table*}

\subsection{Blockchain with SDN}
\begin{figure}[h]
    \centering
    \includegraphics[scale=0.38]{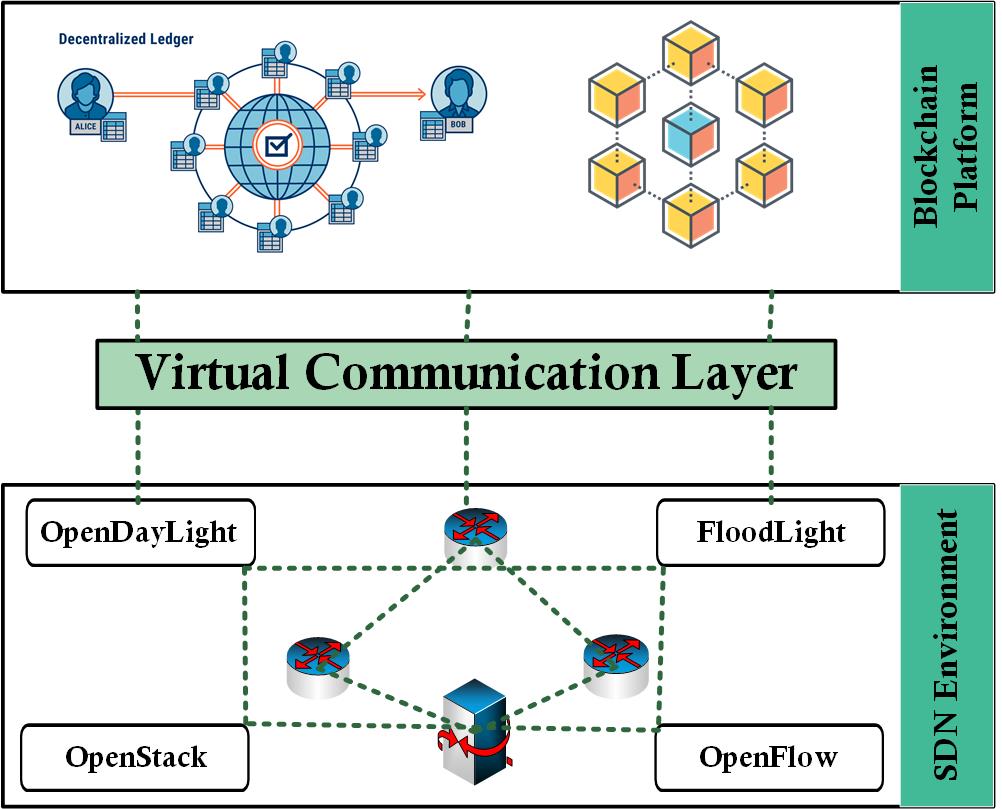}
   \caption{An architecture combining SDN with Blockchain to provide security}
    \label{fig:bss}
\end{figure}

\textcolor{black}{IoT sensor data can be transformed from the sensor level to the SDN environment through an SDN-intelligent gateway. SDN filters the data and Blockchain, and, manages all filtered data efficiently. Further,a virtual communication layer is used to build a relationship between the SDN environment and the Blockchain approach. Also, this layer offers the virtualization technology \cite{hsieh2018mobile} to establish an effective bonding with two methods, as depicted in Fig. \ref{fig:bss}.
}
In the context of IoT architecture, Sharma et al. presented the DistBlockNet paradigm \cite{sharma2017distblocknet}. For different networking technology like SDN and Blockchain, this platform allows for two major rewards. The authors proposed a strategy to update the flow rule table applying the Blockchain technique and formalized the flow rule table. They also measured the results according to different metrics where the analysis presented a better outcome relative to the surviving piece. 
Rahman et al. proposed the \enquote{DistBlockSDN} architecture with network function virtualization (NFV) implementation for a smart city \cite{9290627}. Meanwhile, the authors leveraged a Blockchain approach to achieve high security and privacy. 
Moreover, they also presented an algorithm that can pick the cluster head (CH) consuming low energy. Finally, the authors evaluated the performances of the network using parameters like throughput and packet arrival rate. 
In another study, Navid et al. \cite{navid2019SDIoBoT} presented a novel model to address IoT challenges by combining SDN and Blockchain technologies for future 5G telecommunication networks.

\subsection{Blockchain based Security for Cloud Computing}
\begin{figure}[h]
    \centering
    \includegraphics[scale=0.3]{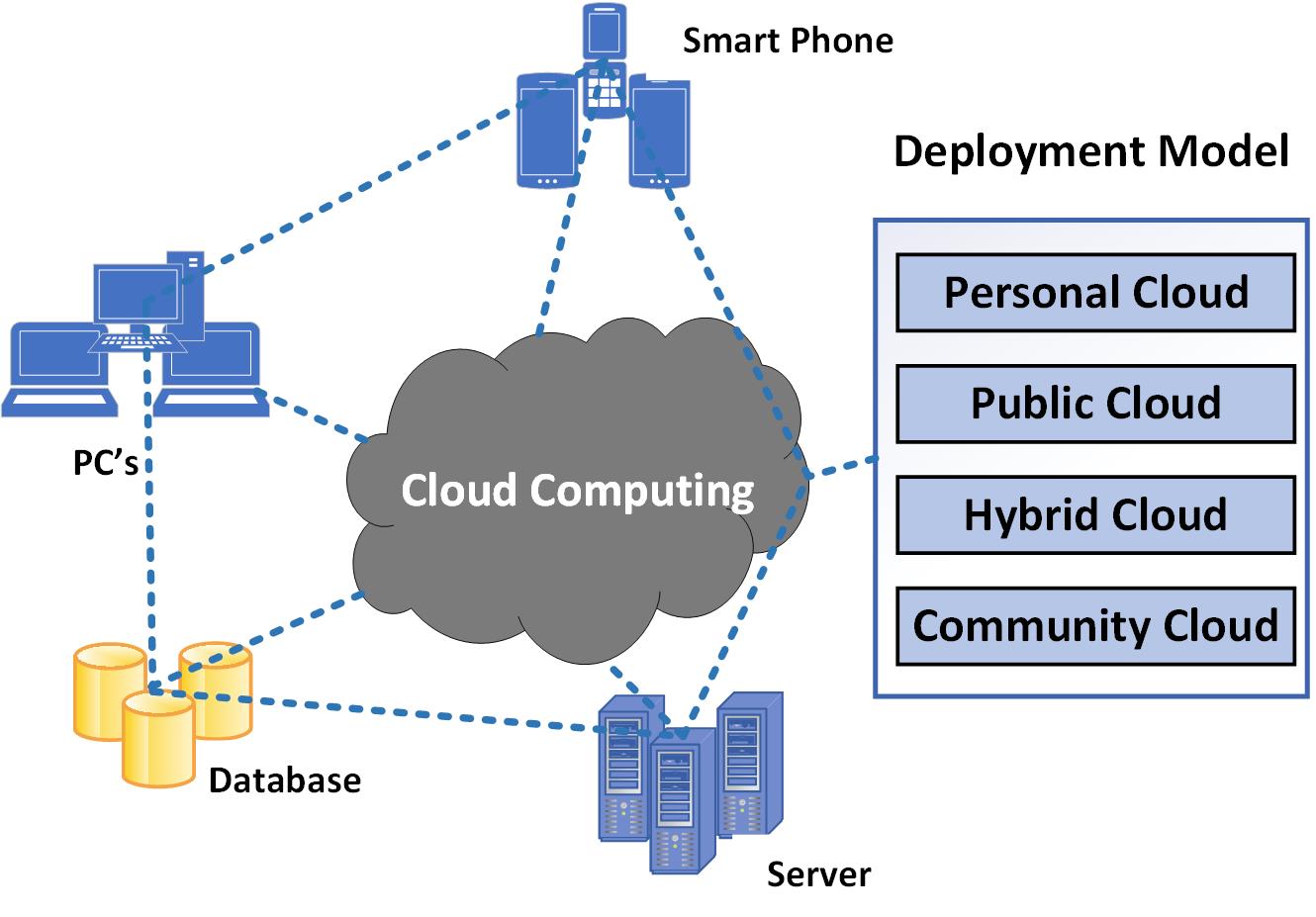}
   \caption{A notional architecture of Cloud computing}
    \label{fig:ccm}
\end{figure}

\textcolor{black}{
Cloud computing is a model that allows for the remote delivery of hardware, software, storage, and other resources via the Internet as services. Different implementation models have been developed according to the application scenario and business purpose; for example to restrict access to the cloud resources only to the employees of an enterprise. There can be personal, public, hybrid, and community deployment models, as reported in Fig. \ref{fig:ccm}.}

Gaetani et al. first presented some research questions based on Blockchain for cloud computing \cite{gaetani2017blockchain}. They presented accurate, high-level solutions to these questions for the European project SUNFISH. 
In related research, Park et al. established the notion of Blockchain technology,and some possible technological cloud computing directions \cite{park2017blockchain}. Also, they presented the high-level security way to cloud computing in various parameters based on the Blockchain concept. 
In this article \cite{rehman2019cloud}, Blockchain technology was used to provide various security services to IoT forwarding devices. They then discussed cloud computing and edge transparent computing technologies in detail. Furthermore, they precisely acknowledged Blockchain measures to safeguard IoT networks from unauthorized threats.
In similar research \cite{sharma2017software}, Sharma et al. introduced a new Blockchain-based disseminated cloud platform with Software-Defined Networking (SDN) enabled controller fog nodes at the network's edge. Moreover, they proposed an excellent combination of fog computing, SDN, and Blockchain.
Furthermore, the authors presented an architecture designed to support high availability, real-time data collection, enhanced scalability, security, and resiliency while keeping low latency.
After that, they also evaluated  parameters like throughput, response time, and accuracy in detecting real-time attacks.\vspace{2mm}

Table \ref{tab:related_work} summarizes the literature review that we have conducted, considering the key technologies employed by each work  and the issue that the authors try to address in a specific application. Moreover, where Blockchain is leveraged we also report the considered implementation platform. \textcolor{black}{After discussing different technologies, we intend to aggregate SDN and Blockchain to secure the cloud services. Since the SDN renders flexibility to the infrastructures and Blockchain provides confidentiality of the services.}

\section{Proposed \enquote{DistB-SDCloud} Architecture for Cloud Computing}
\begin{figure}[ht!]
\centerline{\includegraphics[scale=0.28]{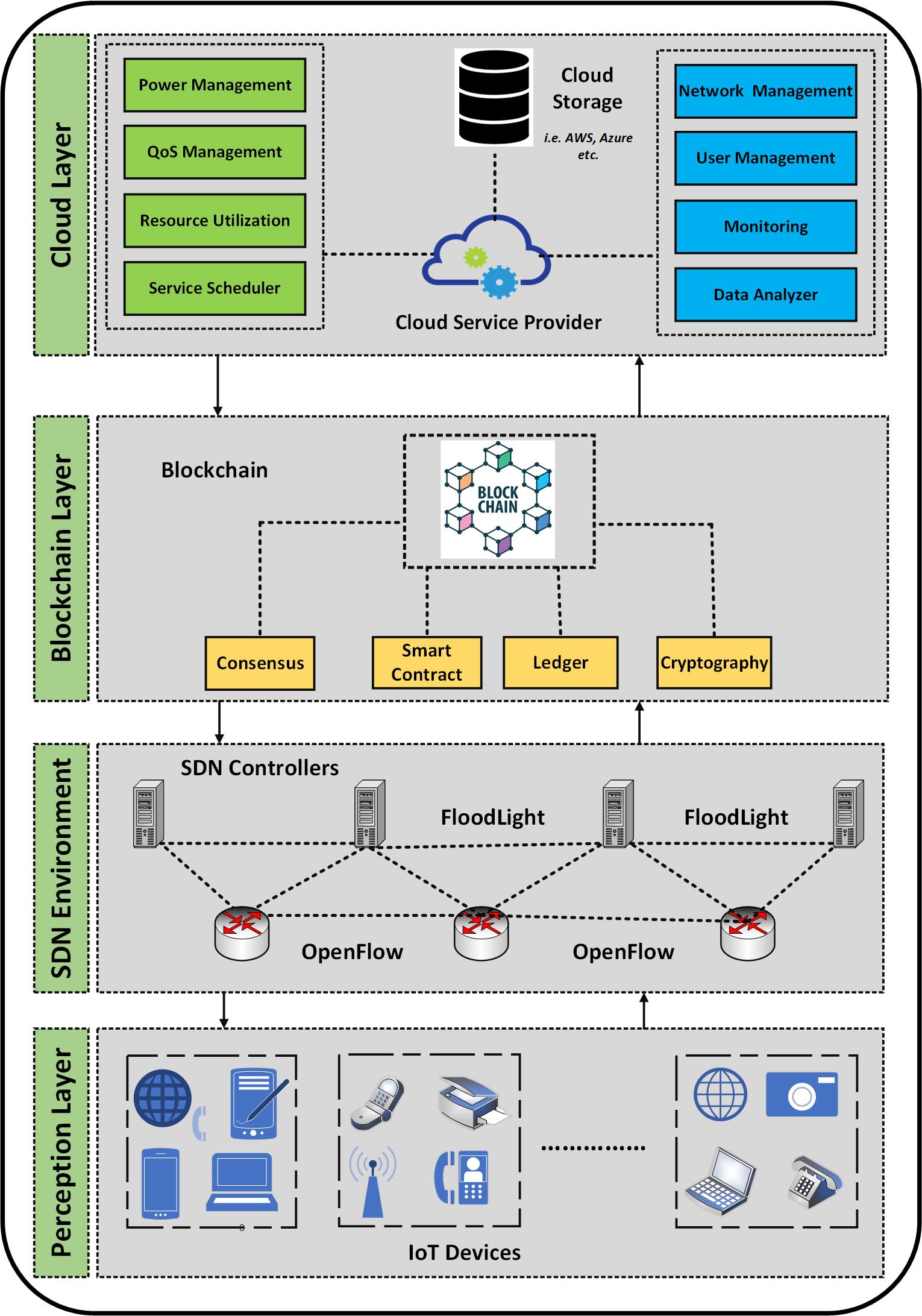}}
\caption{Proposed \enquote{DistB-SDCloud} architecture}    
\label{fig:DisBlockCloud}
\end{figure}

To provide individual security and reliability to cloud applications, we propose a distributed, reliable, Blockchain-based architecture framework that runs these tasks efficiently as shown in Fig. \ref{fig:DisBlockCloud}.
The presented architecture has been divided into four distinct layers. These layers are : Data Extraction, SDN Environment, Distributed Secure Blockchain Methodology, and Cloud Computing Management and Services.

\subsection{Data Extraction}
Several smart sensor-based devices (e.g., an IoT forwarding host) can send sensor data through SDN-enabled gateways controllers, such as firewalls, switches, routers, and various types of data storage devices. These sensor data can be securely used in an SDN-based Blockchain network for various operations \cite{rahman2019distblocksdn, rahman2021study}. Furthermore, sensor data contributes to efficient performance in the distributed architecture.The SDN can then be separated into several planes using sensor data, including data, control, and implementation planes.
The data plane operations regard the collection of sensor data using  multiple controllers, protocols and platforms such as OpenDayLight, OpenFlow and OpenStack respectively, in an SDN environment \cite{MegyesiBAPM17, rahman2020distb}. This layer allows the efficient collection of sensor data using SDN.   

\subsubsection{Data Plane}
\begin{figure}[h]
    \centering
    \includegraphics[scale=0.60]{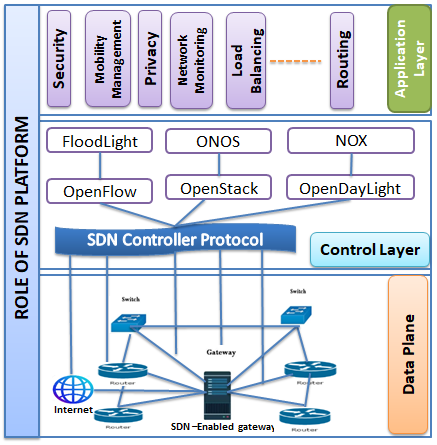}
   \caption{Layered based SDN performance}
    \label{fig:sdne}
\end{figure}

As shown in Fig.\ref{fig:sdne}, the data layer is the lowest in the SDN architecture. 
This plane enables the SDN gateway to communicate properly with sensor-based devices (base station, switches, firewalls, data transfer, and so on). It offers two types of switches virtual and software-based switches, which are commonly used with the Linux operating system. Another type is physical switches, which are related to hardware-based switches and take advantage of the higher flow of physical forwarding devices in the SDN infrastructure plane.
These switches are responsible for forwarding and exchanging packets in network-based applications \cite{xie2018survey}.
In our architecture, the data plane collects the sensor data moving to and from the cloud computing environment.

\begin{figure*}[t]
    \centering
    \includegraphics[scale=0.31]{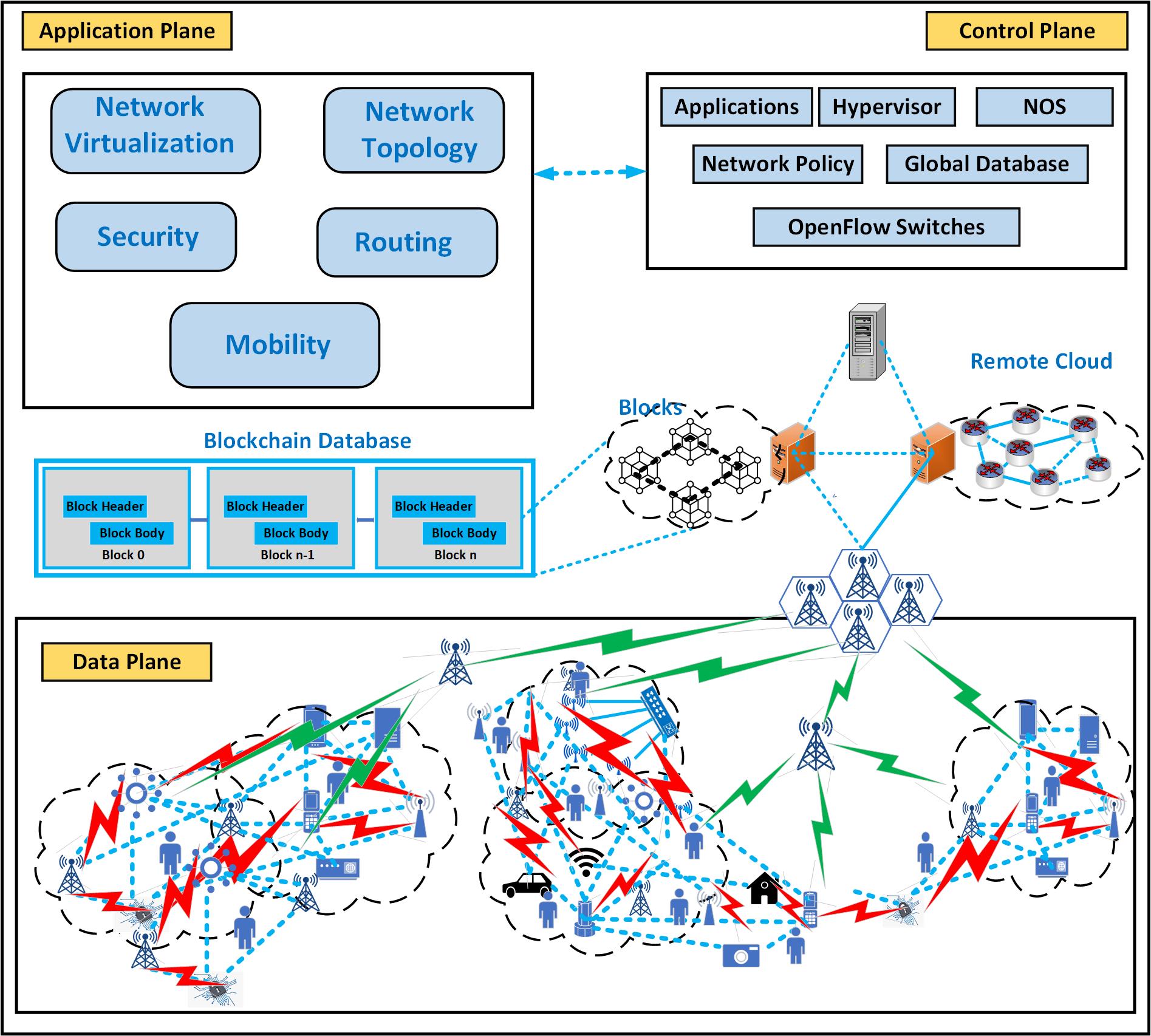}
   \caption{\textcolor{black}{Blockchain and SDN Convergence}}
    \label{fig:BCSDNCon}
\end{figure*}

\subsubsection{Control Plane}
The control plane is the core foundation of the SDN architecture and the foundation of network layer communication.
This plane provides a high-level abstraction, giving the applications  a logically centralized view, although there are multiple distributed controllers at the physical level.
Also, this plane is responsible for the interaction between infrastructure and applications plane in the SDN architecture. It therefore allows numerous networking services for the desired platform.  
The functionalities of this plane are provided through several controllers such as POX, Floodlight, OpenDayLight, Openstack, and Beacon \cite{medved2014opendaylight}, and leveraging protocols like OpenFlow.
The interaction with different devices is realized through different interfaces-  southbound, northbound, and east/west-bound.
The southbound interface communicates with forwarding devices, while the northbound interface is used by application fields; the east and westbound interfaces, communicate with distributed controllers.  
Finally, this controller provides a highly flexible and stable architecture for data to flow into the cloud computing platform. 

\subsubsection{Application Layer}
The service layer is the topmost layer of the SDN. While the lower layers of the SDN-based scheme enable an effective dynamic update of forwarding flow rules, the application layer increases network infrastructure between the control and implementation platforms via virtual or physical forwarding elements. Leveraging the functionalities of the lower layers, the applications can realize complex network configuration and management, network data analytics, or specialized functions targeting specific scenarios such as large data centers. This layer allows services such as smart optimization, mobility management, load balancing, routing, switching, reliability, and network monitoring in the cloud networks, as shown in Fig. \ref{fig:sdne}.

\subsection{Distributed Secure Blockchain-SDN Methodology}
A Blockchain \cite{rahman2020distblockbuilding, rahmanintelligent} is a form of ledger or data system that enables many functionalities to be added to a distributed or decentralized, temperature-resistant facility, as depicted in Fig.\ref{fig:BCSDNCon}. It is based on the identity node of the miner and the request node of the general member. Simultaneously, Blockchain can give effective access control and security to the system design. In essence, Blockchain serves as a secure ledger for recording transactions. It does not store all user activity in a central storage or database.
Further, each user utilizes the same storage at the user end.  Also, they keep all transaction activity and updated copes at the same place to ensure the consensus system.
In the Blockchain environment, every block is can accurately deal with multiple transactions.
Furthermore, every block is also grouped by a hash chain and contains detailed information such as a timestamp, records, existing hash, past data, and non-conflicting transactions.
Based on this concept, we believe that Blockchain is an appropriate technology for ensuring access control in the presented cloud environment.
This approach is organized as following a section name security and access policy \cite{rahman2021smartblock}.

\subsubsection{Security and Access Policy}
By using a Blockchain approach, we provide adequate security to our desired system model.  
This model aims to prevent access of undesired users or unauthorized third parties performing access control,and  deal with data tampering  and loss.
Moreover, the user sends a request to the Blockchain-based server; then, this server takes the necessary action based on the request. At the same time, it will be checked in the server database. If this request is valid, the server sends a positive response to the user to access the various services from the desired server; otherwise, after checking that if the request is not valid, it will be discarded. 

\subsubsection{\color{black}Blockchain and SDN Convergence}
{\color{black}This segment discusses the relationship between Blockchain and SDN which is shown in Fig. \ref{fig:BCSDNCon}. At first, the data layer is responsible for adequately forwarding the intelligent device information to another layer. Moreover, the data passing gateway holds all data to provide the secured platform. Basically, the authors set a confidential platform, Blockchain. This block helps to proffer the temporary database for processing the node data before entering into the cloud applications.
Further, these cloud services can be managed by the desired user remotely. The whole process is executed by the SDN Platform efficiently. SDN uses OpenFlow switches to perform all activities in the desired application. After finishing the convergence between Blockchain and SDN, the user can get the different services such as security, remote routing, mobility, privacy, and virtualization into the cloud data management applications \cite{Chaudhary2019BESTBS}.}

\subsubsection{\color{black}Block Creation and Validation}
\begin{figure}[H]
    \centering
    \includegraphics[scale=0.33]{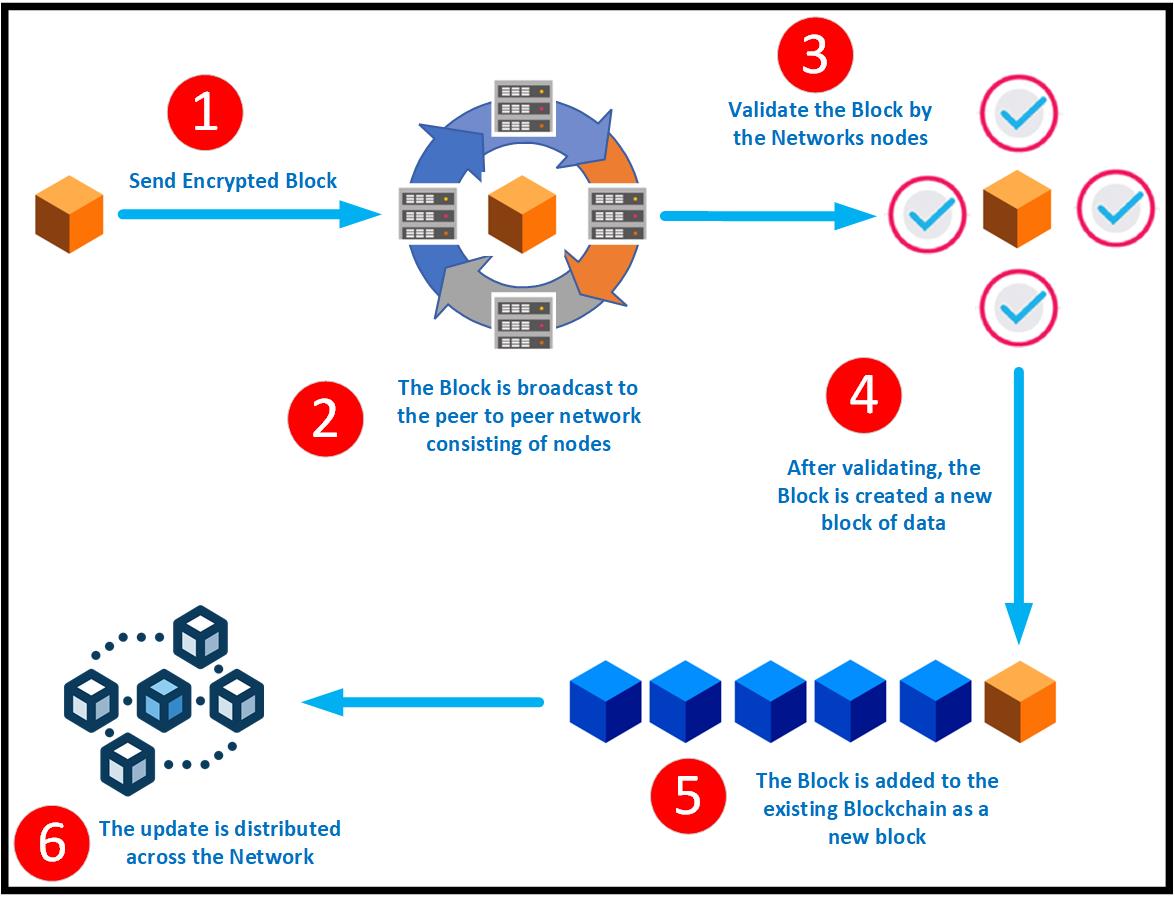}
   \caption{\textcolor{black}{Block Creation and Validation}}
    \label{fig:bcvp}
\end{figure}

{\color{black}To provide security into the presented system, the authors have considered distributed platform Blockchain. The block creation and validation process of Blockchain are shown in Fig. \ref{fig:bcvp}. Firstly, the authors send an encrypted block to the peer-to-peer networking environment. Then, these blocks are validated by networks nodes through the mining process \cite{guo2020blockchain}. After validating the process, the block is created a novel data block. Further, this block is appended to the Blockchain network as a new block. Finally, the distributed ledger is updated itself manner efficiently. } 

\subsection{Attack Mitigation in Cloud Environment Blockchain-based SDN Approach}

SDN is very vulnerable to attack. Some of the most common attacks are DDoS attacks \cite{azadpreventive}, DoS attacks, flooding attacks, and many more 
now that the cloud environment is fundamental to various innovations. Attacks against this cloud storage scenario have become a significant attraction for many intruders. The overall system could be down for some time because of these attacks. We also proposed a method to recognize and block these attacks to address this problem so that the device can operate accordingly. The mechanism of the Blockchain has been used to support the SDN controllers. The controller will detect potential attacks according to this instruction. The second issue is the avoidance or defense after the identification of the intruder. In the Blockchain approach, the data packets are transmitted block by block. In this case, only the authorized blocks can take their place, and unauthorized blocks are discarded from the chain.

\color{black}
\subsection{Cloud Computing Management and Services}
\begin{figure}[h]
    \centering
    \includegraphics[scale=0.27]{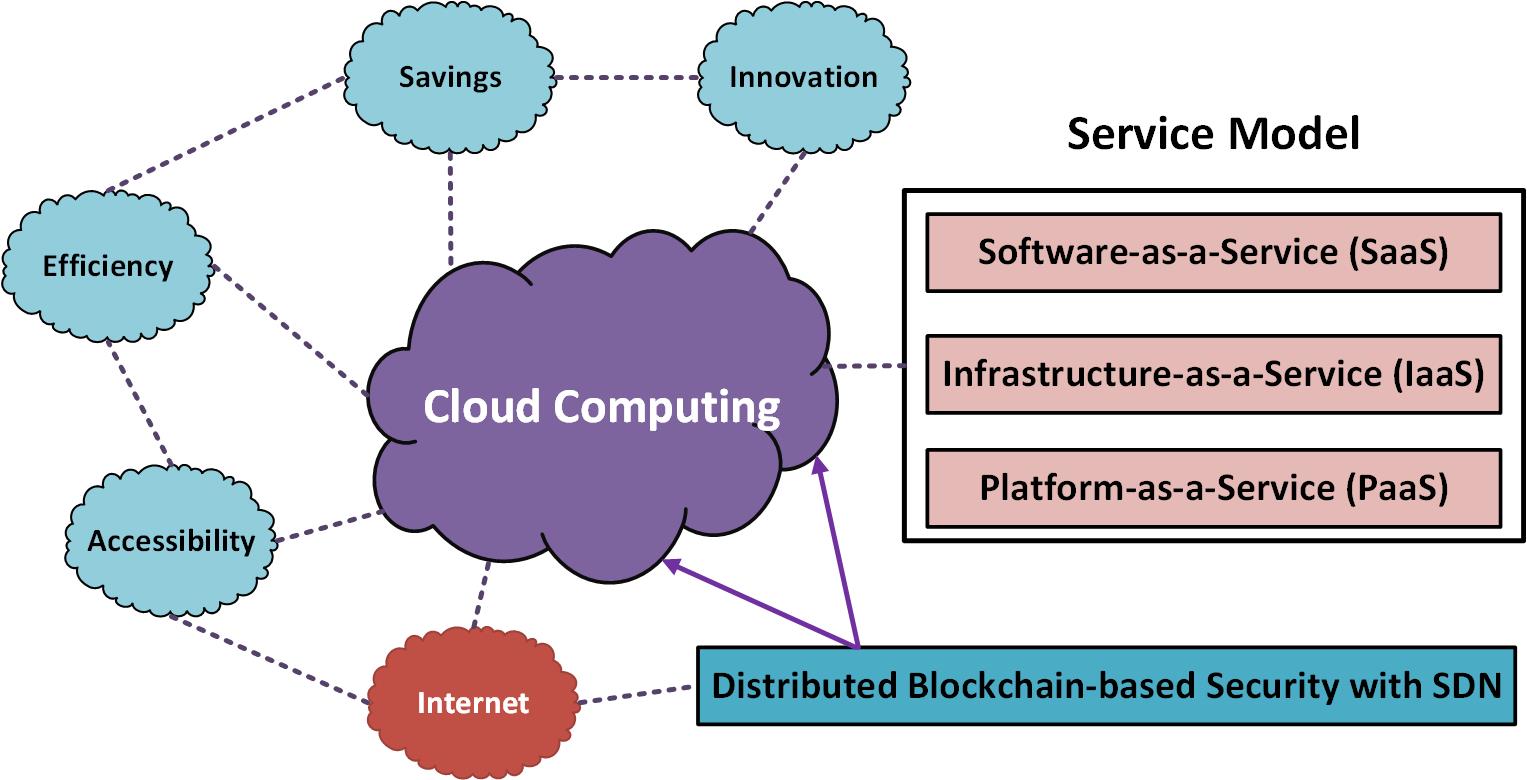}
   \caption{Distributed Blockchain-SDN based Cloud computing services}
    \label{fig:ccs}
\end{figure}

Our proposed model enhances various services based on a distributed Blockchain approach in the cloud computing environment.
This BC-based SDN architecture provides benefits like flexibility, accessibility, security, privacy for retrieving, and storing numerous resources in the cloud computing platform.
Without the involvement of the SDN approach, Blockchain alone \cite{rehman2019cloud} cannot  provide improved reliability, high stability, a logically centralized controller, and an increased load balancing capability in the presented architecture.
While the accessibility and availability of services and resources in the cloud eventually depends on Internet speed, cloud computing resources performs better with as a Service (SaaS), Platform as a Service (PaaS), and Infrastructure as a Service (IaaS) using the presented architecture correctly, as shown in Fig. \ref{fig:ccs}.

\section{Performance Analysis and Discussion}

\subsection{Performance Measurement Parameters}
To evaluate the performance of the proposed architecture, we have considered two main parameters, namely throughput and communication overhead. 

We calculate the throughput using  equation Eq.(1).  
\begin{equation}
Throughput=\frac{\sum_{n=i}^{n}CBRrece}{Simulation\ Time}
\end{equation}
CBR is the abbreviation for Constant Bit Rate.

We also use equation, Eq.(2) to calculate the communication overhead.   
\begin{equation}
Communication\ Overhead=\frac{\sum RTRPacket}{\sum CBRrece}
\end{equation}
RTR indicates the Response to Request time. 

\subsection{Environment Setup}
To create the environment in which our proposed system will operate, we have used Mininet as network emulator, together with the Mininet-Wifi module to emulate a wireless network configuration.
OpenStack was used as a cloud storage platform, while OpenFlow was chosen as the protocol within the SDN architecture.
Our tests were carried out on a computer with the Ubuntu (Linux) operating system, a Core(TM)-i7 processor, 3.40GHz CPU, and 500GB of hard drive space.
Furthermore, we have also used the Wireshark to sniff traffic flowing into the network to test the effectiveness of our SDN-based Blockchain network. 
Table \ref{tab:simEnv} summarizes the parameters we have used in our simulation environment.

\begin{table}[H]
\caption{Simulation Environment}
\vspace{2mm}
\centering
\begin{tabular}{p{1.3cm} p{3cm} p{3cm}}
\hline
 & \textbf{Parameters Name} & \textbf{Values}\\
\hline
\multirow{2}{4em}{General Parameters} &  Packet Analyzer & Wireshark \\ 
& Platform for cloud storage & OpenStack \\

\hline
\multirow{4}{4em}{SDN Parameters} & SDN Controllers & 5\\
& OpenFlow switches & 4\\
& Gateways & 2 \\
& SDN Routing Protocol & OpenFlow\\
\hline
\multirow{4}{4em}{Blockchain Parameters} & Blockchain platform & Ethereum\\
& Number of transactions & Variable\\
& Block header & 80 bytes\\
& Proof type & Proof of Work (PoW)\\
\hline

 \multirow{7}{4em}{Others Parameters} & Simulation Area & 1000m X 1000m \\
& Number of IoT devices &  1-50  \\
& Simulation Times & 500s \\
& Data Rate & 12 Mbps\\
& Initial Energy Values of IoT devices & 12-15 j\\
& Initial Trust value & 5 j\\
& Node Transmit Packet Size & 100-512 bytes \\

\hline
 
\end{tabular}
\label{tab:simEnv}
\end{table}

\subsection{Performance Evaluation of Blockchain with SDN Controller in Cloud Computing Environment}
\begin{figure}[H]
    \centering
    \includegraphics[width=0.98\linewidth]{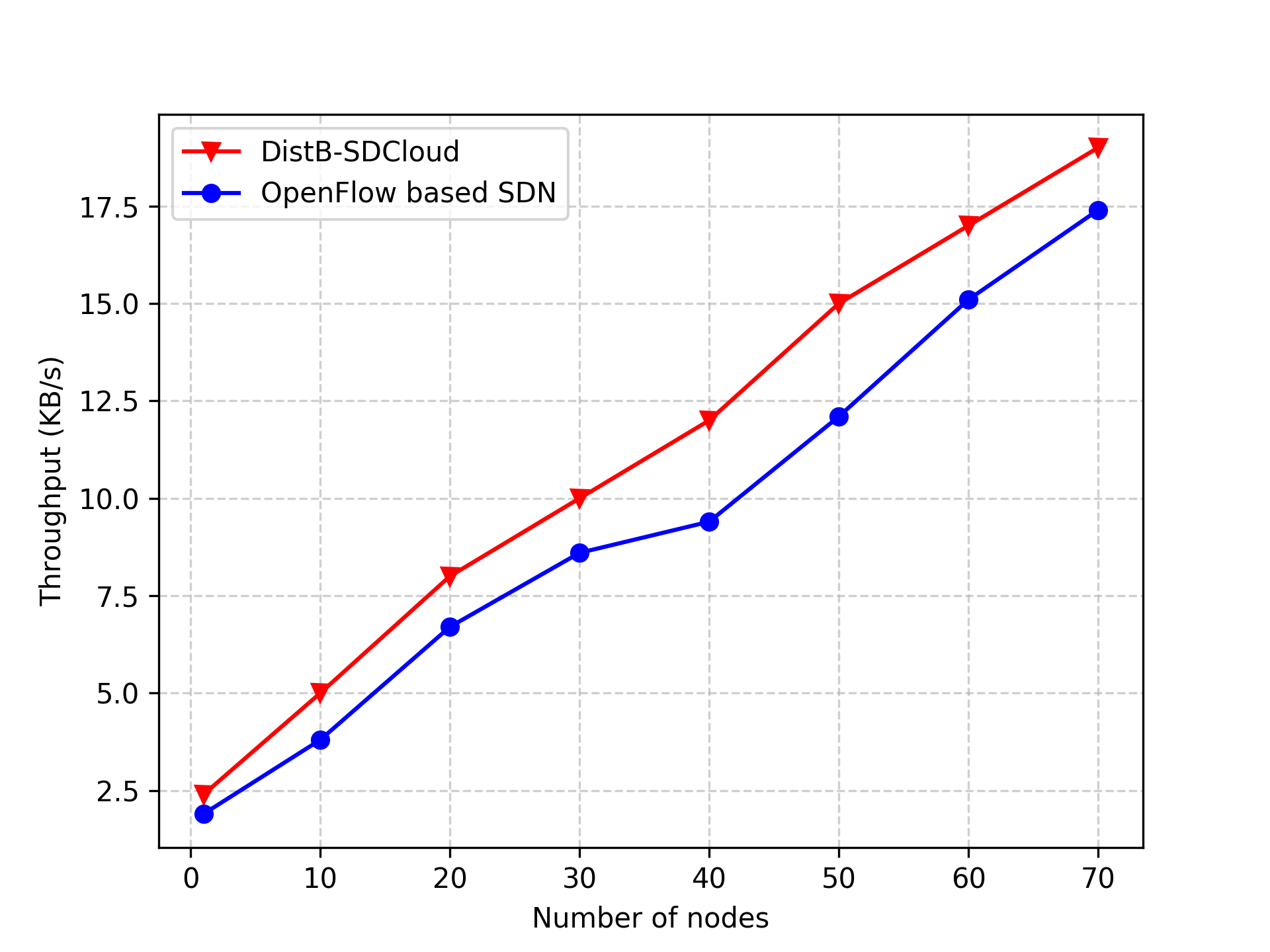}
    \caption{Throughput comparisons according to the number of nodes in the network.}
    \label{fig:thc}
\end{figure}

\textbf{Throughput:}
Throughout this section, we have assessed the performance of our suggested model in terms of many metrics such as throughput, packet arrival rate, and file transfer operation.
First, as shown in Fig. \ref{fig:thc}, we calculated throughput using the number of nodes. Meanwhile, throughput assessments between OpenFlow-based SDN and our proposed architecture \enquote{DistB-SDCloud} are being conducted and are graphically depicted in this figure.
Furthermore, we have noticed that when the number of nodes is reduced, the throughput is nearly the same. However, as the number of nodes increases, the throughput increases as well.
Finally, we compared our proposed framework \enquote{DistB-SDCloud} to one that uses only SDN and OpenFlow, and found that our suggested scheme outperforms the other. \vspace{2mm}

\begin{figure}[h]
    \centering
    \includegraphics[width=0.98\linewidth]{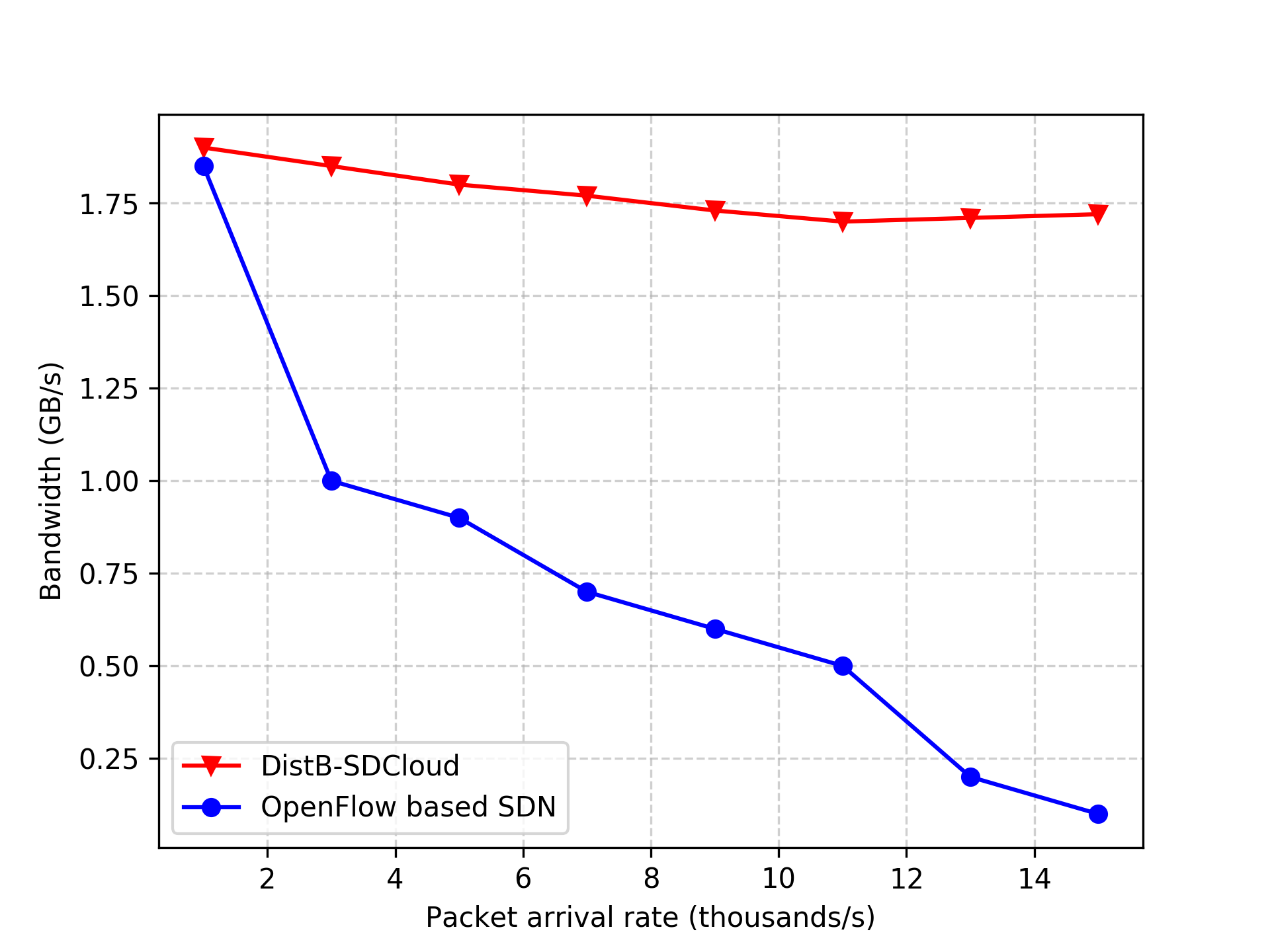}
   \caption{Bandwidth measured varying on the packet arrival rate.}
    \label{fig:pac}
\end{figure}

\textbf{Packet Analysis:}
Fig. \ref{fig:pac} depicts the system performance when dealing with an increasing number of packets. To this goal, this figure reports the bandwidth (GB/s) versus the current packet arrival rate (thousand/s), comparing the results of our proposed system versus  an OpenFlow-based SDN.
The tested packet rates range from $190$ to more than $1400$ packets per second for both tested models.
From Fig. \ref{fig:pac}, we can see that when the packet rate increase (which could be, for example, a clue to a network attack being underway), the bandwidth is dramatically decreased in the OpenFlow-based SDN. 
On the contrary, the performance of our presented model stays unaltered, even when increasing the attack rate, and even with the highest tested packet rate, proving its robustness against suddenly increased loads, caused by malicious or intended activities. 

\begin{figure}[h]
    \centering
    \includegraphics[width=0.99\linewidth]{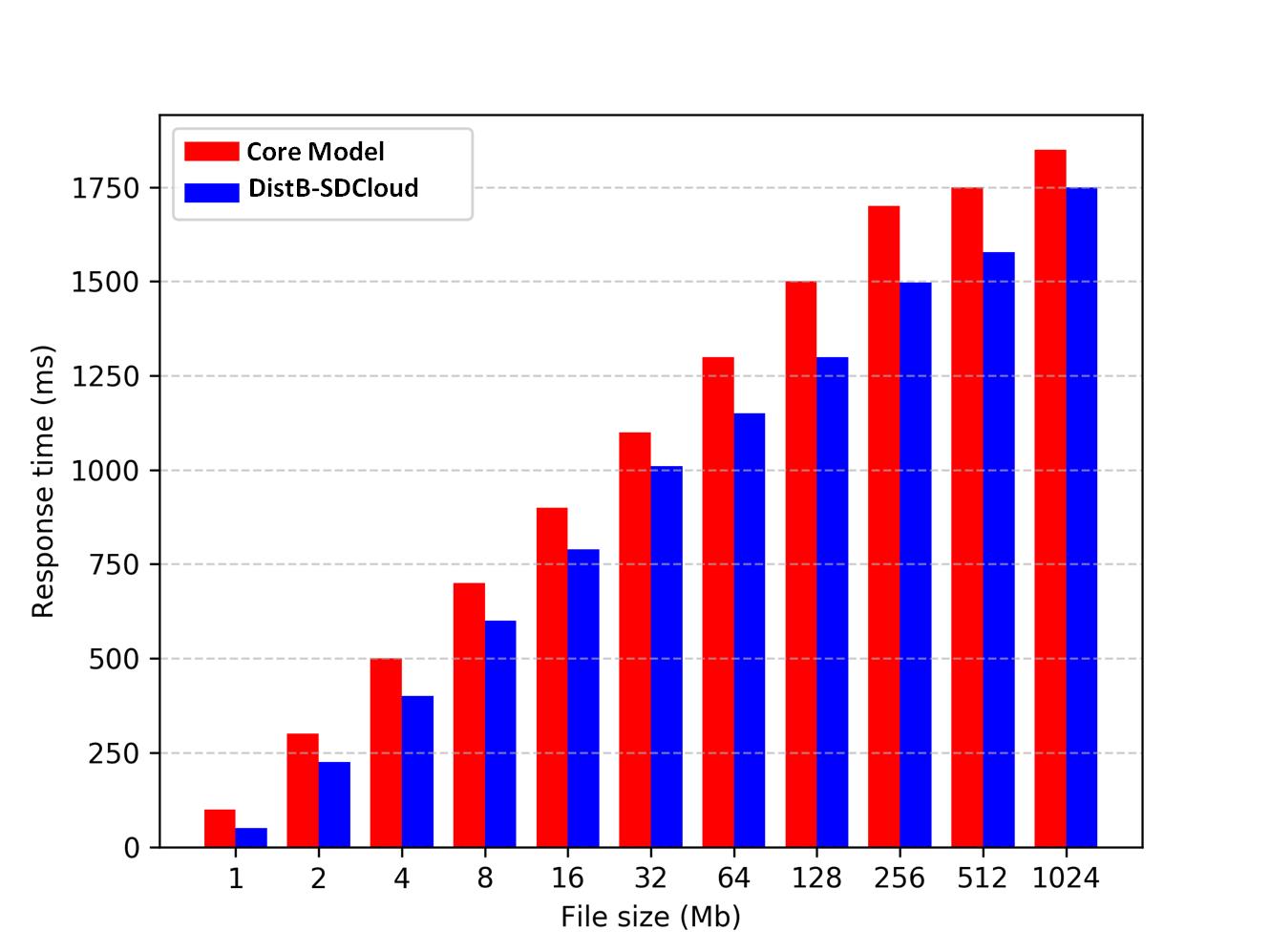}
   \caption{Response time for variable-sized file transfers.}
    \label{fig:FO}
\end{figure}

\textbf{Response Time:}
Fig. \ref{fig:FO} shows the performance of file operations for the core and  presented models.
This figure reports the response time for file transfer operations when varying the file sizes.
Indeed, when increasing the file size, the response time  also increases.
However, the proposed model constantly reports a lower and therefore better response time performance compared to the core model. 
Moreover, we also observed that our model is could achieve considerable file sizes compared to the existing core-based system.

\begin{figure}[h]
    \centering
    \includegraphics[scale=0.33]{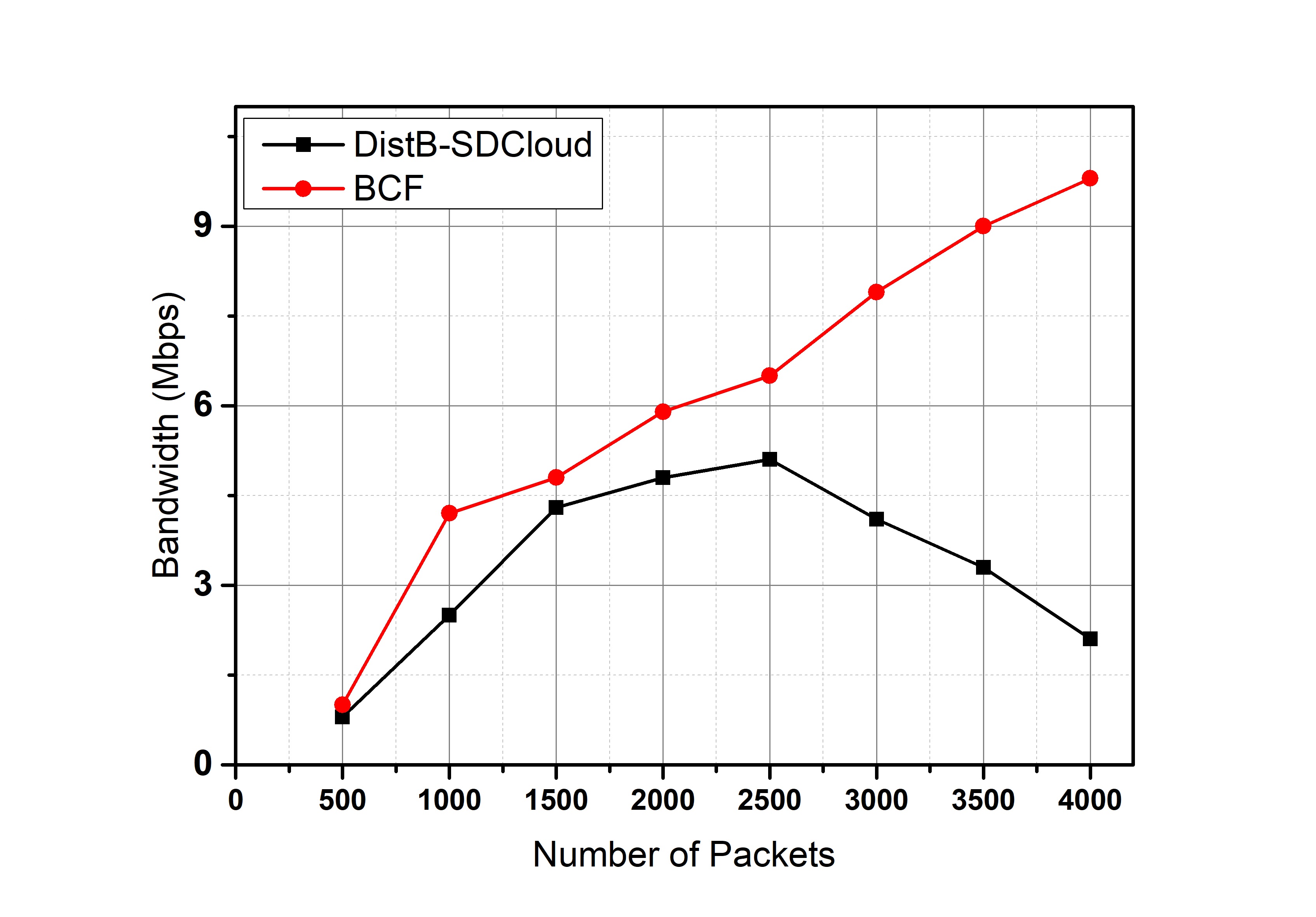}
   \caption{\textcolor{black}{Bandwidth Vs. Number of packets}}
    \label{fig:BNP}
\end{figure}


\begin{figure}[h]
    \centering
    \includegraphics[scale=0.33]{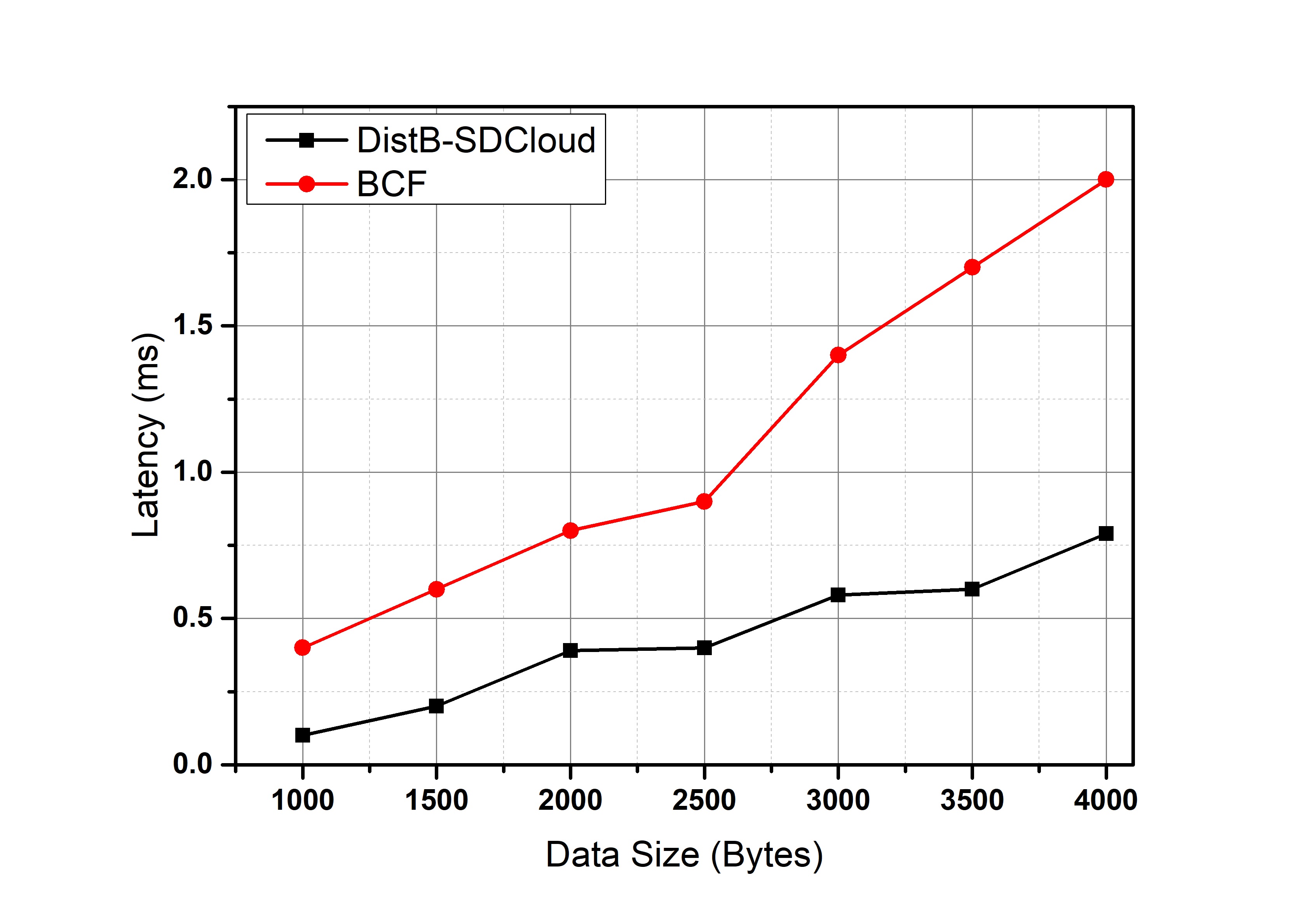}
   \caption{\textcolor{black}{Latency Vs. Data size (Bytes)}}
    \label{fig:Latency}
\end{figure}

\textbf{Bandwidth and Latency Analysis:}
{\color{black}We analyze another performance parameters Bandwidth with respect to the number of packets. Here, According to the analysis of Yazdinejad et al., we have measured our performance of bandwidth and delay based on the BCF \cite{yazdinejad2020energy} method. In Fig. \ref{fig:BNP} shows the outcomes between the proposed system \enquote{DistB-SDCloud} and BCF model. Moreover, the authors considered SDN controllers and SDN-based OpenFlow protocol for creating the desired network. Then, they applied protocol-based rules to measure bandwidth concerning several packets. Initially, both models show the performance is almost nearest in the networking environment, which was exactly the same when the packet number was 500. After increasing the number of packets, the presented model shows better performance compared with the BCF model. For the attachment of SDN-controllers- the system model achieves scalability, reliability, network stability; that's why this shows the best performance than the BCF model. The proposed framework shows the highest performance when the packet number is 2500 and the rest of the packets. On the other hand, Fig. \ref{fig:Latency} presented the latency based on the data size. Due to the effect of SDN (switch, controllers) involvement, the proposed system \enquote{DistB-SDCloud} can be capable of responding fast than the BCF model.}

\subsection{Performance Analysis in Different Attacks Environment}
\begin{figure}[h]
    \centering
    \includegraphics[width=0.98\linewidth]{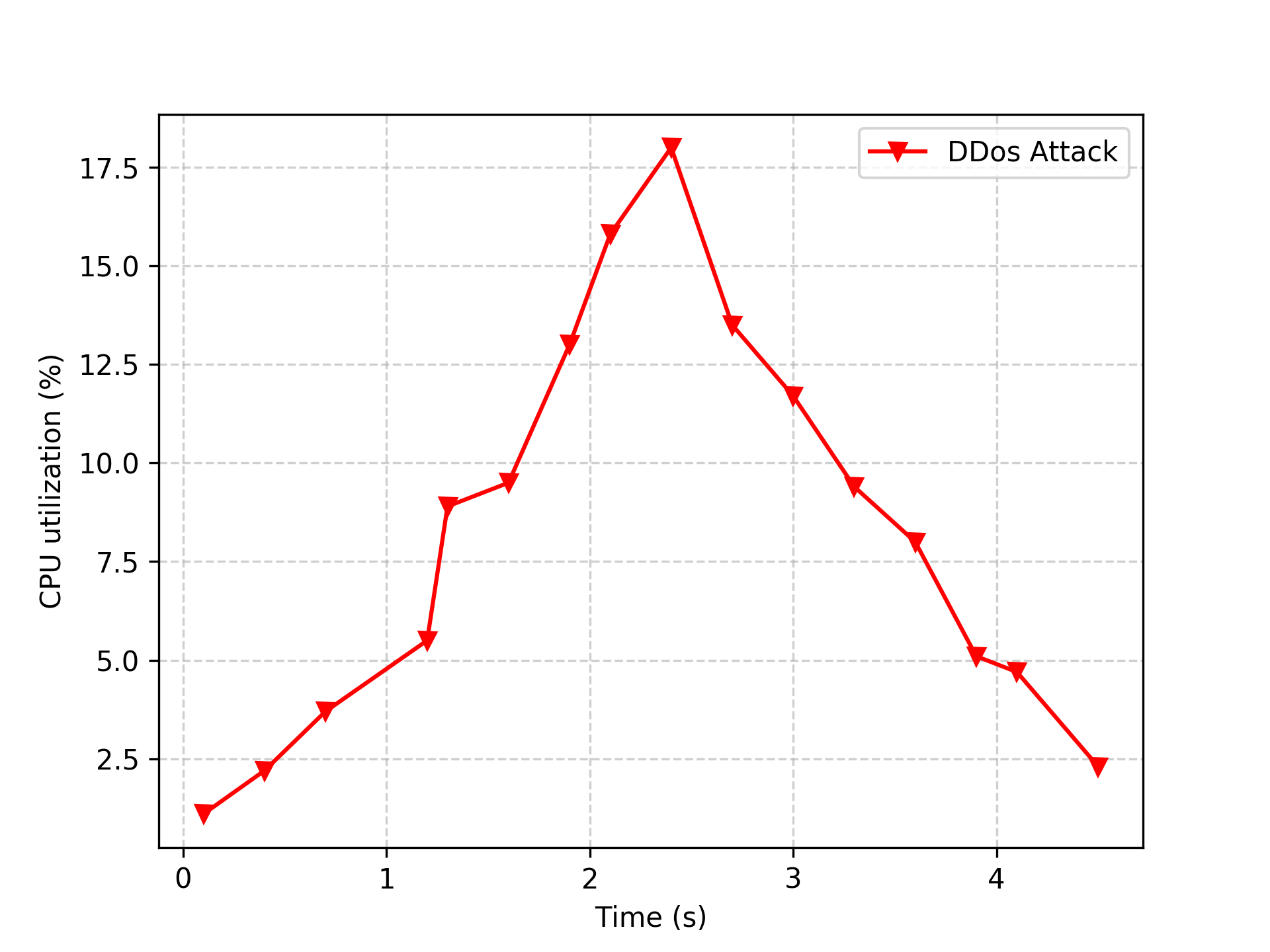}
   \caption{CPU utilization during DDoS attacks}
    \label{fig:dda}
\end{figure}

After comparing throughput, packet arrival rate, and reaction time, we discovered that our suggested model outperforms an OpenFlow-based SDN model in terms of throughput, packet arrival rate, and reaction time. 
In this section, we also discuss the complexity in terms of CPU utilization of our architecture during network attacks. In Fig. \ref{fig:dda} shows the analysis of CPU utilization for distributed Denial-of-Service (DDoS) assaults on our infrastructure when several services are running in the background. Furthermore, we employed a learning set to record CPU consumption during a DDoS attack.This figure also illustrates the average CPU consumption in different apps based on the \enquote{DistB-SDCloud} scheme when DDoS assaults are undertaken.
The simulated attack began at a value of about $1.1$ value, and the attack rate increased over time.
When we look at the evolution of CPU use over time, we can see how it rises for a while, then swiftly drops to lower levels, demonstrating that \enquote{DistB-SDCloud} can provide a sound defense against this assault.

\subsection{Discussion}
This study, combines  of two leading technologies, SDN and Blockchain, to achieve maximum efficiency and security in a cloud computing infrastructure.
One essential finding is that using Blockchain with SDN  improves security and increases the overall throughput, while keeping adequate response times and CPU utilization under control, especially in the context of network attacks.

\section{Future Scopes}
\subsection{Artificial Intelligence (AI) in SDN-IoT Networks}
\textcolor{black}{
The functionality of an SDN is to control a system programmatically, hence AI and machine learning models could be deployed to the controllers from where various decisions could be taken within a system. The flow management of different SDN based IoT networks using AI for cloud \cite{manogaran2021ai}. This technology is also is in action in the field of 5G enabled by SDN and Network Virtualization \cite{lin2021toward}. The scalability of the systems based on these technologies is another concerning issue. In this case AI can also play a role \cite{belgaum2021role}. Moreover, SDN-IoT networks could be smart and self-propelled with the help of AI methods.
}

\subsection{AI with Blockchain}
\textcolor{black}{It is hard to find any research area where AI is not contributing. Most systems are being automated with the applications of AI. As Blockchain can work with the security and confidentiality of data, it could be combined with AI to construct smart systems with better security. They can offer solutions to  modern problems including medical issues \cite{nguyen2021blockchain}, business revolution \cite{sharma2021iot}, e-learning \cite{hung2021ai} etc. Conceptual modeling in governance is also being encountered with the analysis of AI and Blockchain \cite{alshamsi2021artificial}. Digital marketing could be another prominent field where these technologies can be used.
}

\subsection{Security and Privacy Issues in IoT-AI}
\textcolor{black}{Privacy and security some of the most popular area of IoT research.  AI is an advanced technology that can detect a security breach automatically in smart cities \cite{lv2021ai} and smart healthcare based on IoT networks \cite{gopalan2021iot, jobaersecure2021}. Identifying and resolving different intruders using AI models can lead a system to a successful real-life application. Some researchers have started to apply AI in  security issues for 6G networks. The difficulties and opportunities, for example, are discussed by Siriwardhana et al. \cite{siriwardhana2021ai}. 
}

\vspace{2mm}
There are countless applications for utilizing these technologies to improve security and intelligence in a system. Aside from cloud computing, the combination might provide numerous benefits in various industries ranging from smart cities to healthcare to business and privacy.

\section{Conclusion}
Recently, the demand for cloud computing services has been rising rapidly, with the number of  customers increasing dramatically.
However, cloud computing is faces several threats and challenges, like security, privacy, compliance, compatibility, control, and reliability. 
To address these threats, researchers have proposed several tips and techniques; however, the security of these services still faces several  challenges that remain an open research question. 
In this work, we propose the \enquote{DistB-SDCloud} architecture to enhance the security and confidentiality of cloud computing systems.
We have leveraged a distributed Blockchain approach integrated into an SDN architecture to increase the security, privacy, stability, reliability, successful accessibility, and the confidentiality of the cloud computing services for users. As part of future research, we intend to effectively implement this architectural approach in diverse applications such as the fog and edge computing domains. The proposed system paradigm will then be more reliant to by SDN and Blockchain technologies.

\EOD

\end{document}